\documentclass{aa} 
\usepackage{txfonts}
\usepackage[latin1]{inputenc}  
\usepackage{graphicx}

\begin{document}
\title{Photodissociation of organic molecules\\
 in star-forming regions\\ \vspace{0.5cm}
 \Large{III. Methanol}}
\author{S. Pilling\inst{1} \and R. Neves\inst{2} \and A. C. F. Santos\inst{3} \and H. M. Boechat-Roberty\inst{2}}
\institute{Laboratório Nacional de Luz Síncrotron, Caixa Postal
6192, CEP 13084-971, Campinas, SP, Brazil \and Observatório do
Valongo, Universidade Federal do Rio de Janeiro, Ladeira Pedro
Antônio 43, CEP 20080-090, Rio de Janeiro, RJ, Brazil. \and
Instituto de Física, Universidade Federal do Rio de Janeiro, Caixa
Postal 68528, CEP 21941-972, Rio de Janeiro, RJ, Brazil}
\offprints{S. Pilling,\\ \email{spilling@lnls.br}}
\date{Received / Accepted}
%
\abstract{ The presence of methyl alcohol or methanol (CH$_3$OH) in several
astrophysical environments has been characterized by its high abundance that depends on
both the production rate and the destruction rate. In the present work, the
photoionization and photodissociation processes of methanol have been experimentally
studied, employing soft X-ray photons (100-310 eV) from a toroidal grating monochromator
(TGM) beamline of the Brazilian Synchrotron Light Laboratory (LNLS). Mass spectra were
obtained using the photoelectron photoion coincidence (PEPICO) method. Kinetic energy
distribution and abundances for each ionic fragment have been obtained from the analysis
of the corresponding peak shapes in the mass spectra. Absolute photoionization and
photodissociation cross sections were also determined. We have found, among the channels
leading to ionization, about 11-16\% of CH$_3$OH survive the soft X-rays photons. This
behavior, together with an efficient formation pathways, may be associated with the high
column density observed in star-forming regions. The three main photodissociation
pathways are represented by COH$^+$ (or HCO$^+$) ion release (with ejection of H$_2$ +
H), the dissociation via C-O bond rupture (with strong charge retention preferentially
on the methyl fragment) and the ejection of a single energetic (2-4 eV) proton. Since
methanol is very abundant in star forming regions, the produced protons could be an
alternative route to molecular hydrogenation or a trigger for secondary dissociation
processes or even to promote extra heating of the environment.

}

\titlerunning{Photodissociation of organic molecules in SFRs III: Methanol}
\authorrunning{Pilling et al.}
\maketitle

\section{Introduction}

Methyl alcohol or methanol (CH$_3$OH), the simplest alcohol, is one of the most abundant
molecules detected toward various astrophysical regions, including hot molecular cores
(HMCs) associated with low- and high-mass star-forming regions (Requena-Torres et al.
2006, Remijan et al. 2004, Pei et al. 2000; De Buizer et al. 2000; Menten 1991; Norris
et al. 1993, Caswell et al. 1993 and references therein); dense molecular clouds
(Irvine, Goldsmith \& Hjalmarson 1987; Tielens \& Alamandola 1987), circumstellar and
proto-stellar regions (Schutte et al. 1999; Keane et al. 2001; Goldsmith et al. 1999;
Pontoppidan et al. 2003, 2004; Grim et al. 1991) and in comets (Mumma et al. 2001;
Bockelee-Morvan et al. 1994; Crovisier \& Bockelée-Morvan 1999 and references therein).
In these objects, the radiation field can drive several photophysical and photochemical
processes, including molecular photodissociation. The products of organic molecule
dissociation (e.g. reactive ions and radicals) can fuel the formation of interstellar
complex molecules such as long carbon chain molecules.

The maser emission of CH$_3$OH in the GHz range is its main fingerprint and, to date, at
least six catalogues have presented this spectroscopic signature in star forming regions
(e.g. Schutte et al. 1993; Caswell et al. 1995; Walsh et al. 1997; Slysh et al. 1999;
Szymczak, Hrynek, \& Kus 2000). The X-ray photons from star-forming regions are capable
of traversing large column densities of gas before being absorbed. X-ray-dominated
regions (XDRs) in the interface between the ionized gas and the self-shielded neutral
layers could influence the selective heating of the molecular gas. The complexity of
these regions possibly allows a combination of different scenarios and excitation
mechanisms to coexist together (Goicoechea et al. 2004).

The formation mechanisms for interstellar methanol occur both in the gas phase and on
grain surfaces. In dense clouds, methanol is frozen onto grains. The enhanced abundance
of methanol in the  warm gas has been taken as evidence that the methanol is released
from grains in these regions.

Both the UMIST (Millar et al. 1991) and Standard Model (Lee, Bettens \& Herbst 1996)
reaction data set include a particularly simple gas-phase chemistry for methanol
formation directly via radiative associative reaction of a methyl radical and water
molecules followed by electronic recombination with hydrogen liberation
\begin{equation}
CH_3 + H_2O \longrightarrow CH_3OH_2^+ + h\nu
\stackrel{e^-}{\longrightarrow} CH_3OH + H,
\end{equation}
However, as pointed by Millar, Herbst \& Charnley (1991) the gas phase reactions would
not be enough to justify the relative methanol abundances detected, of about 10$^{-7}$
or more, and grain mantle involvement was required.

From the theoretical models of Tielens \& Whittet (1997) and Charnley, Tielens \&
Rodgers (1997) the successive hydrogenation of CO in interstellar ices
 \footnotesize %
\begin{eqnarray}
\nonumber 
CO \stackrel{H}{\longrightarrow} HCO \stackrel{H}{\longrightarrow}
H_2CO \stackrel{H}{\longrightarrow}
CH_3O \texttt{ and/or } CH_2OH \stackrel{H}{\longrightarrow} CH_3OH \\
\end{eqnarray}
 \normalsize %
has been proposed to produce molecules like methanol. Sorrel (2001) has also proposed
that accretion of gas-phase H, O, OH, H$_2$O, CH$_4$, NH$_3$ and CO onto dust grains
sets up a carbon-oxygen-nitrogen chemistry in the grain that can also produce methanol.
As a consequence, a high concentration of free OH, CH$_3$ and NH$_2$ radicals is created
in the grain mantle mainly by photolysis reactions. Once these radicals are created,
they remain frozen in position until the grains heat up. As this occurs, the radicals
become mobile and undergo chemical reactions among themselves and with other adsorbed
molecules to produce complex organic molecules including CH$_3$OH.

Over the years, many experimental studies involving UV photolysis and proton bombardment
on ices were tested and some of these routes have been fully accepted, Shalabiea \&
Greemberg (1994) have proposed a route for formation of solid methanol on grain mantles
from UV photolysis of formaldehyde-water ices,
\begin{equation}
H_2CO + H_2O + h\nu \stackrel{ }{\longrightarrow} CH_3OH + O,
\end{equation}
They predict a methanol photoproduction rate per incident photon of about 1.6 $\times
10^{-3}$ photon$^{-1}$.

On the other hand, Allamandola, Sandford \& Valero (1988) and Shuttle et al. (1996) have
performed UV photolysis of H$_2$O+CO ice but the methanol yields are insufficient to
explain the observed abundances in interstellar ices.

Another set of reaction pathways was studied by Moore \& Hudson (1998) and Hudson \&
Moore (1999) who employed proton bombardment on CO + H$_2$O and CH$_4$+ H$_2$O ices,
\begin{equation}
CO + H_2O \stackrel{protons}{\longrightarrow} CH_3OH + O,
\end{equation}
and
\begin{equation}
CH_4 + H_2O \stackrel{protons}{\longrightarrow} CH_3OH + H_2.
\end{equation}
The results have shown that the methanol yield on both carbon monoxide-water and
methane-water ices, despite the evident production of methanol, could not justify alone
the detected CH$_3$OH abundance in interstellar ices. However, perhaps due to the high
abundance of interstellar carbon monoxide-water ices, this process may dominates the
interstellar methanol production. In summary, some of the above mentioned methanol
formation routes seem able to account for the observed abundances.

Watanabe \& Kouchi (2002), also using hydrogen bombardment on
CO+H$_2$O ices, have found a methanol yield up to 17\% for a proton
dose of about  $4 \times 10^{18}$ H$^+$/cm$^{2}$, revealing a large
correlation of methanol production with the proton flux.

The photodissociation of methanol in the gas phase has been studied experimentally and
theoretically in the ultraviolet region by several authors (Shi et al. 2002; Tang et al.
2002, Zavilopulo 2005). However, despite some photoabsorption studies in the X-ray range
(Ishii \& Hitchcook 1988; Prince et al. 2003; Burton et al. 1992) there are few studies
focusing on the photodissociation pathways in this photon energy range. Azuma et al.
(2005) have investigated the state-selective dissociation processes of the O1s
core-excited methanol and deuterated methanol. Stolte et al. (2002) were able to
experimentally isolate a specific bond-breaking process in methanol (CH$_3$OH) by
observing the anion fragment, OH$^-$.

The present work aims to examine the photoionization and photodissociation of gaseous
methanol by soft X-rays, from 100 eV up to 310 eV, including the energies around the
carbon K edge (C1s edge) at $\sim$ 290 eV. In section 2, we present briefly the
experimental setup. The results of the photoionization and photodissociation of the
methanol molecule in the presence of soft X-ray photons, together with the main
photodissociation pathways and the determination of the absolute cross sections are
presented and discussed in section 3. Finally, in section 4, final remarks and
conclusions are given.

\section{Experimental}

The experiment was performed at the Brazilian Synchrotron Light Laboratory (LNLS), in
Campinas, São Paulo, Brazil. The experimental setup has been described in detail by
(Boechat-Roberty, Pilling \& Santos, 2005 - PAPER I; Pilling, Boechat-Roberty \& Santos,
2006 - PAPER II). Briefly, soft X-ray photons in the energy range of 100 to 310 eV, from
a toroidal grating monochromator (TGM) beamline, perpendicularly intersect the effusive
gaseous sample inside a high vacuum chamber. The ionized (cation) recoil fragments
produced by the interaction with the radiation were accelerated and detected by two
micro-channel plate detectors, after mass-to-charge (m/q) analyzing by a time-of-flight
mass spectrometer.  Conventional time-of-flight mass spectra (TOF-MS) were obtained
using the correlation between one Photoelectron and a Photoion Coincidence, PEPICO,
described by Santos et al. (2001), Boechat-Roberty et al. (2005) and Pilling et al
(2006). Negative ions may also be produced and detected, but the corresponding
cross-sections are negligible.

The base pressure in the vacuum chamber was in the $10^{-8}$ Torr range. During the
experiment the chamber pressure was maintained below $10^{-5}$ Torr. The pressure at the
interaction region (volume defined by the gas beam and the photon beam intersection) was
estimated to be $\sim$ 1 Torr (10$^{16}$ mols cm$^{-3}$). The measurements were done at
room temperature. The methanol sample was commercially obtained from Sigma-Aldrich with
purity better than 99.5\%. No further purification was used except for degassing the
liquid sample by multiple freeze-pump-thaw cycles before admitting the vapor into the
chamber.

\section{Results and discussion}

\begin{figure}[t]
\resizebox{\hsize}{!}{\includegraphics{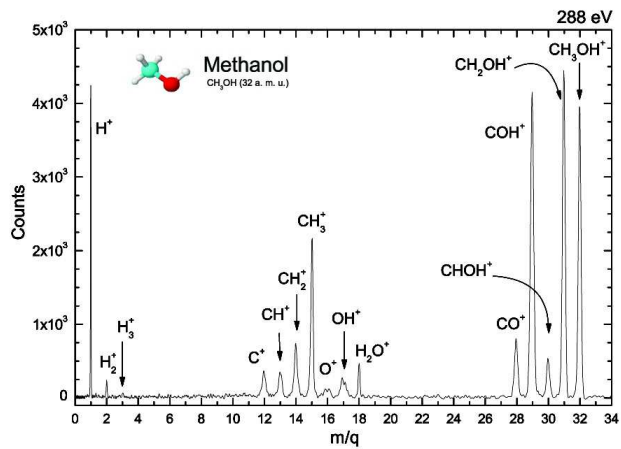}} \caption{Time-of-flight mass
spectrum of methanol after exposure to 288.3 eV X-ray.} \label{fig-ms}
\end{figure}

Figure~\ref{fig-ms} shows the mass spectrum of methanol obtained at the photon energy of
288 eV, over the C1s$\rightarrow 3s$ resonance energy (Prince et al. 2003). We can see,
for example, the methyl group fragments  (mass from 12 to 15 a.m.u) and the CO group (28
to 31 a.m.u). The O group and the recombination water ion (16 to 18 a.m.u) are also
produced. There are also small a mounts of reactive H$_2$ and H$_3$ ions. The
photodissociation pathways for these species become more effective in multiple
ionization processes. These features will be described in detail in a future publication
together with the same behavior in other highly hydrogenated organic molecules. The
parent ion CH$_3$OH$^+$ peak remains with high intensity showing the intrinsic stability
of methanol at this photon energy.

The most produced ions are COH$^+$ (or HCO$^+$), CH$_2$OH$^+$, and H$^+$, followed by
the parent ion and the methyl group.

\begin{figure}[!h]
 \centering
\includegraphics[scale=0.552]{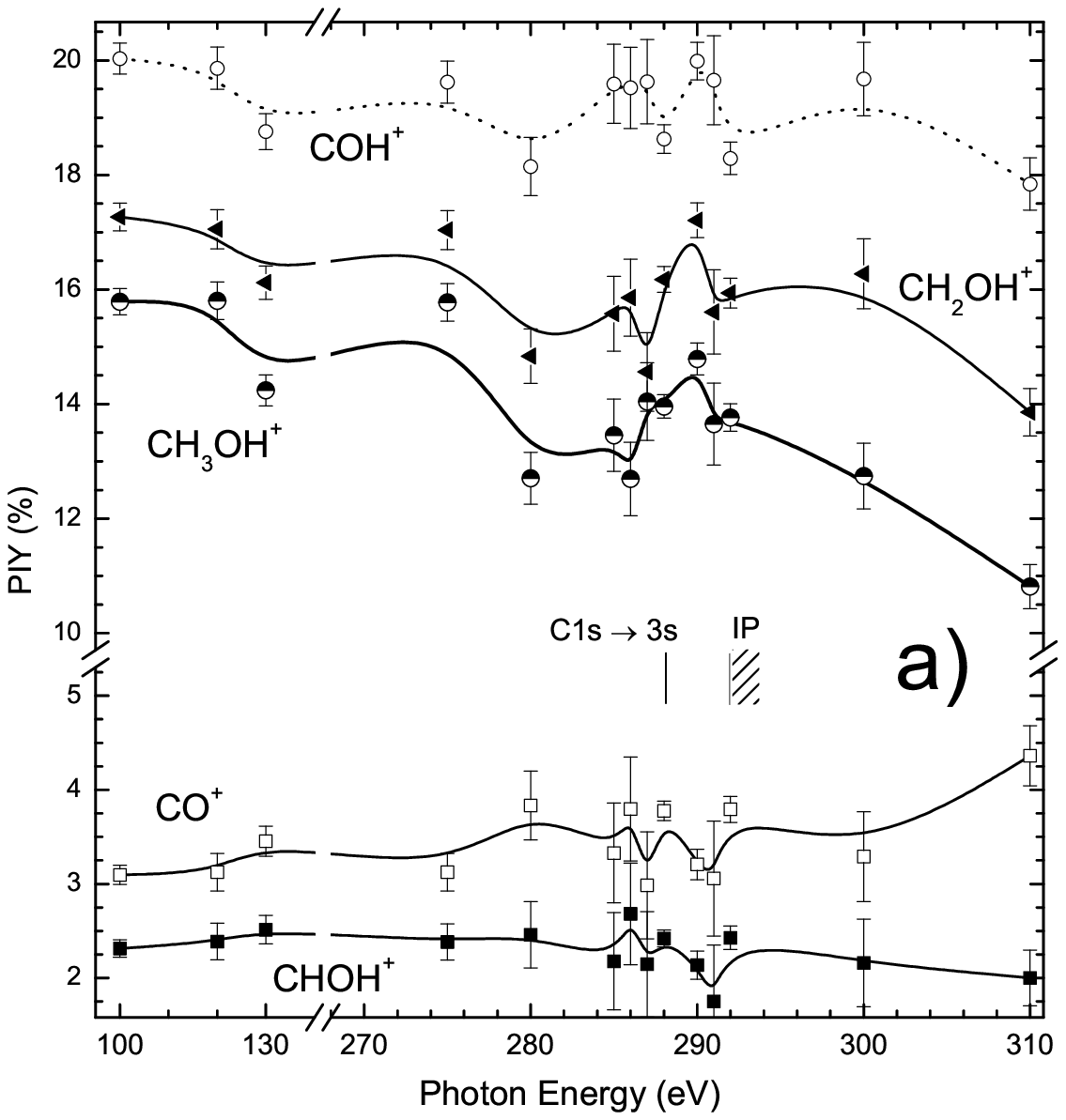}
\includegraphics[scale=0.552]{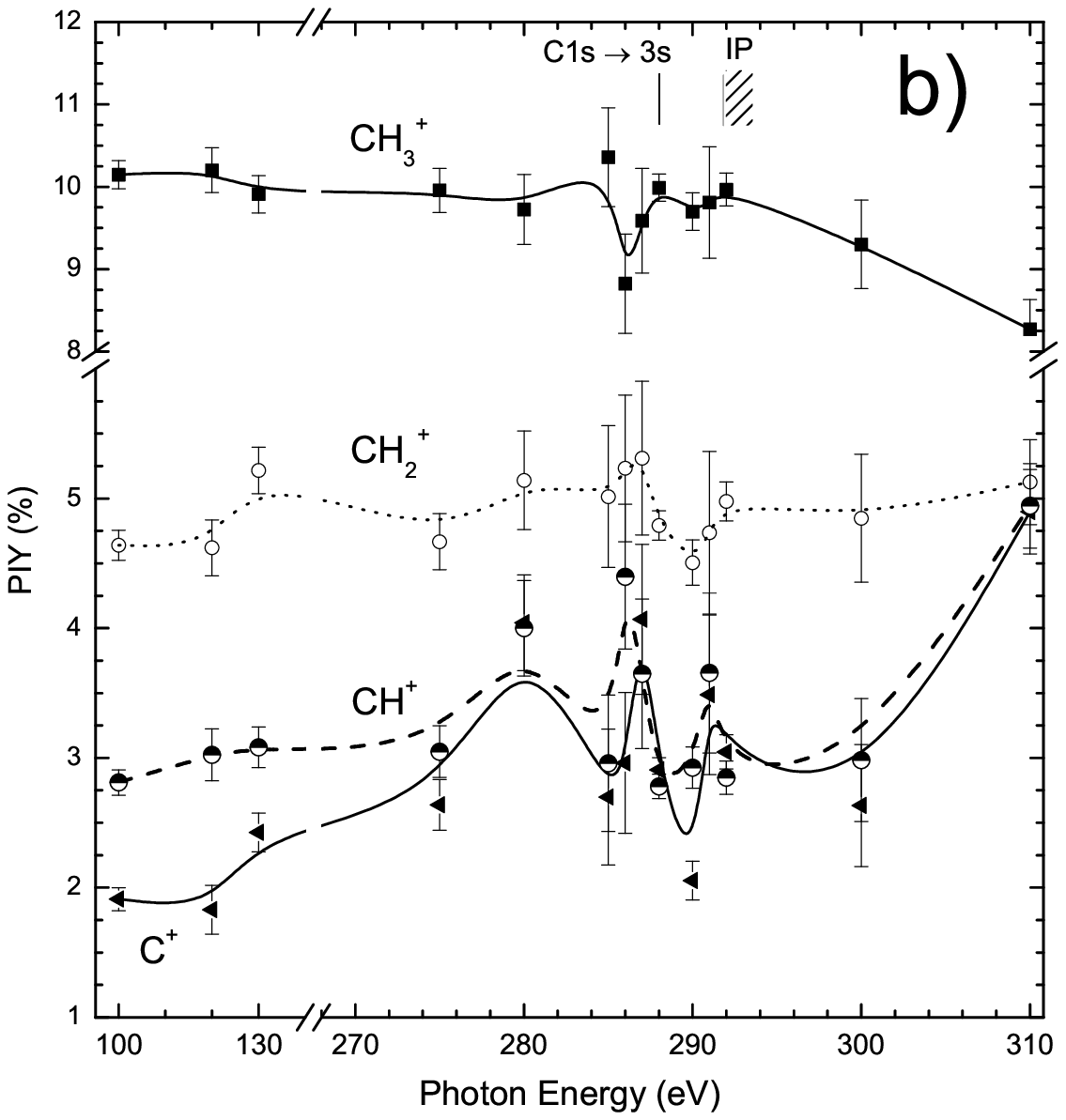}
\includegraphics[scale=0.552]{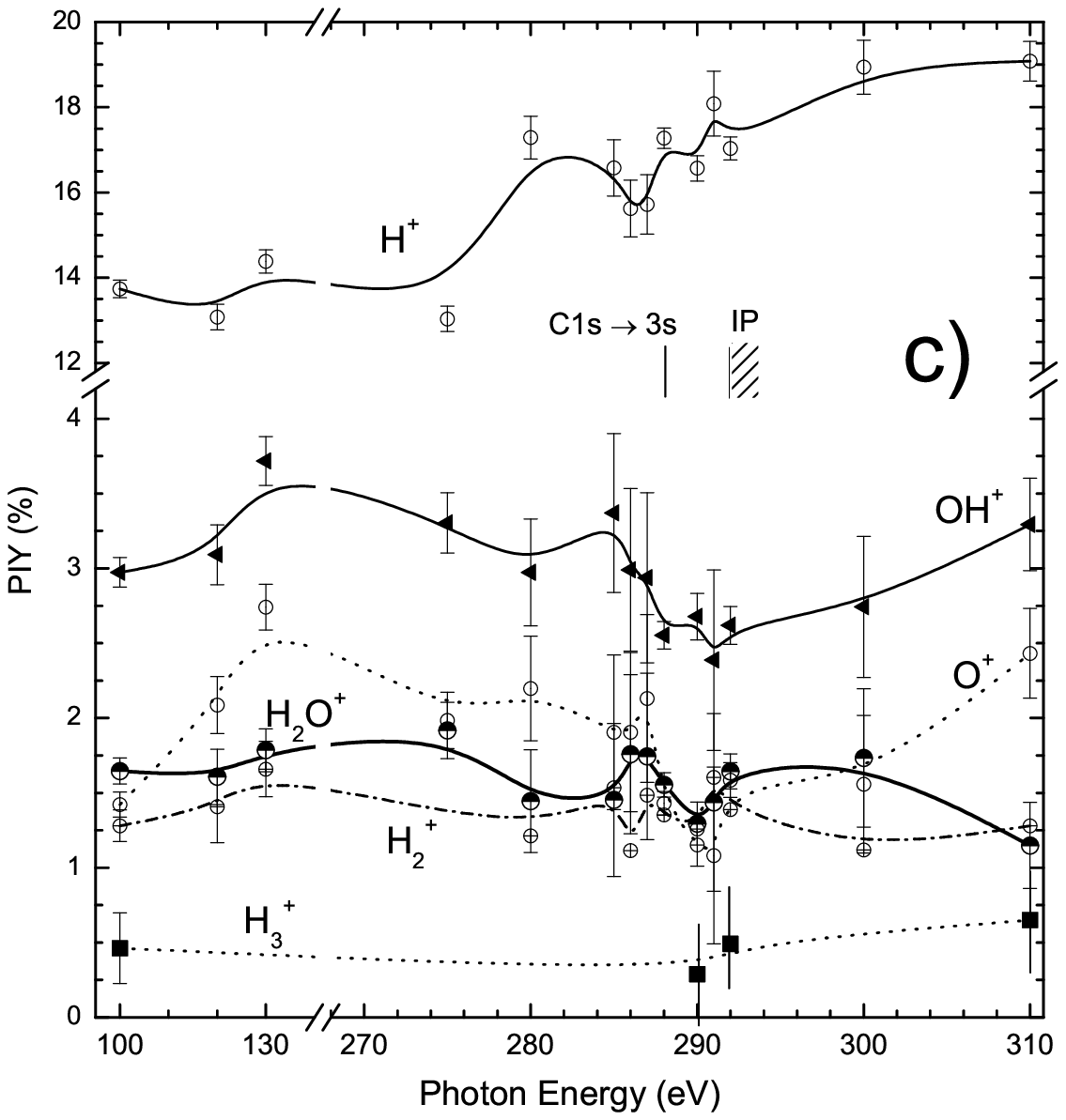}
\caption{Partial ion yield (PIY) of the fragments release by the CH$_3$OH molecule as a
function of photon energy. See details in text.} \label{fig-piy}
\end{figure}

\subsection{Partial ion yield and kinetic energy release}

The partial ion yield (PIY), or relative intensities, for each
fragment $i$ were obtained by
\begin{equation}
PIY_i = \left( \frac{A_i }{A^{+}_t} \pm \frac{ \sqrt{A_i}+ A_i
\times ER/100}{A^{+}_t} \right) \times 100\%
\end{equation}
where $A_i$ is the area of a fragment peak, $A^{+}_t$ is the total area of the PEPICO
spectrum. The error factor $ER$ (2 \%) is the estimated error factor due to the data
treatment.

Figures~\ref{fig-piy}a,~\ref{fig-piy}b and \ref{fig-piy}c show the relative intensities
(PIY) of fragments from methanol photodissociation in the photon energies range of 100
to 310 eV . Figure~\ref{fig-piy}a presents the CO group fragments and also the parent
ion CH$_3$OH$^+$, in Figure~\ref{fig-piy}b and c, we can see the C group fragments and O
group fragments, respectively. In the bottom figure we could also see the PIY of H$^+$,
H$_2^+$ and H$_3^+$. The location of the ionization potential (IP) at 292.32 eV and the
C1s $\rightarrow$ 3s resonance (Prince et al. 2003) are indicated in the figures. The
statistical uncertainties were below 10\%.

A clear bump can be seen in the fractions of all released fragments near the C1s
resonances and the IP energies. The methyl ion as a result of the neutral OH release
represent about 8 to 10\% of the total ion yield. Its counterpart, the hydroxyl ion, as
a result of the neutral fragments liberation, CH$_3$ (or CH$_2$ + H), represent only
about 3 to 4 \% of the total ion yield. This large difference between the yield of these
two fragments, after C-O bond rupture indicates that, during the dissociation process,
there is a significant preference in the charge retention by methyl over the hydroxyl.
This behavior has also been seen in the photodissociation of acetic acid (PAPER II).

\begin{figure}[t]
\resizebox{\hsize}{!}{\includegraphics{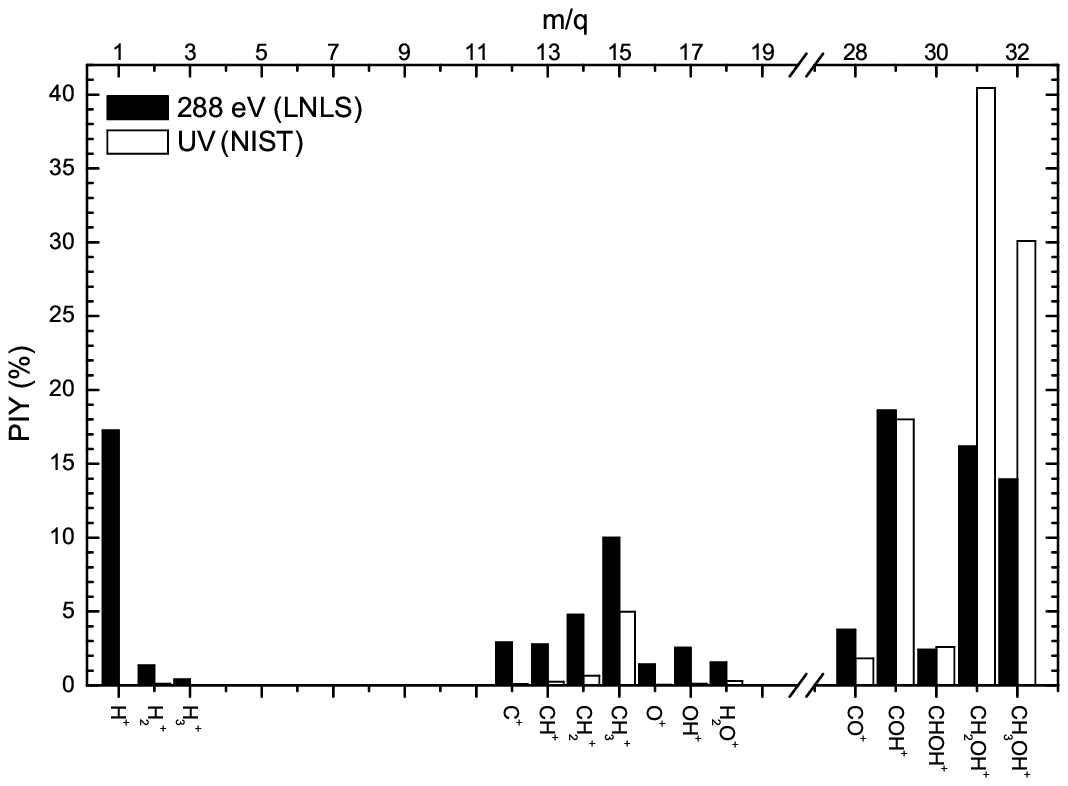}} \caption{Comparison between
partial ion yield (PIY) of methanol fragments in soft X-ray and 70 eV electrons measured
at (NIST). The dissociation induced by 70 eV electrons is very similar to the
dissociation induced by 21.21 eV UV photons.}
 \label{fig-piycomp}
\end{figure}

We present in the Figure~\ref{fig-piycomp}, a comparison between partial ion yield of
the CH$_3$OH fragments by soft X-rays (288 eV) and the fragment yield upon bombardment
with 70 eV electrons measured at National Institute of Standards and Technology (NIST).
The dissociation induced by 70 eV electrons is very similar to the dissociation induced
by 21.21 eV (He I Lamp) photons. In both cases the ionization occurs in the valence
shell. The degree of destruction of the methanol is at least twice as large in the soft
X-ray case than by photons in the UV region. Several fragments present a different
dissociation pattern, as far as PIY are concerned, in X-ray field when compared to the
UV field. As an example, we mention the large enhancement of CH$_2$OH$^+$ produced by UV
radiation with respect to the opposite behavior presented by CH$_3^+$ and all lower mass
ions which seem to be more efficiently produced by X-rays photons. The COH$^+$ (or
HCO$^+$) and CHOH$^+$ (or H$_2$CO$^+$) ion yields seem to present only small changes
with respect to photon energies in the UV-X-ray range.

The absence of more doubly ionized fragments in the PEPICO spectra indicates that doubly
ionized methanol dissociates preferentially via charge separation. The dynamics of the
doubly and triple ionized methanol molecule will be the subject of study of a future
publication.

As pointed out by several authors (Largo 2004, Woon 2002, and references therein)
understanding the pathways of formation of biomolecules present in star-forming regions
and other gaseous-dusty astronomical media is extremely important not only to classify
the chemistry of those regions and, but also, to give some clues about the origin and
possible spread of life in the Universe. Despite the success of \emph{ab initio}
theoretical calculations, the endothermic ion-molecule reactions are not considered as
viable in the interstellar medium and only exothermic reactions have been accepted as
viable mechanisms. However, with the knowledge of the kinetic energy (or at least with
its value range) of some radical and ionic photofragments, some endothermic ion-molecule
reactions could be competitive and, in extreme situations, even become more efficient
than those exothermic reactions.

\begin{table*}
\centering %
\caption{Relative intensities (PIY) and kinetic energy $U_0$ release by fragments in the
methanol mass spectra, as a function of photon energy (at 100, 288, 292 and 310 eV).
Only fragments with intensity $>$ 0.1 \% were tabulated. The estimated experimental
error was below 10\%.} \label{tab:piy}
\begin{tabular}{ l l l r r r r }
\\
\hline \hline
\multicolumn{2}{c}{Fragments}    &  & \multicolumn{4}{c}{PIY (\%) / $U_0$ (eV)}\\
\cline{1-2}  \cline{4-7}
   $m/q$        & Attribution     &  & 100 eV             & 288 eV       & 292 eV         & 310 eV       \\
\hline
1       & H$^+$                  &  & 13.7 / 2.2     & 17.3 / 2.9  & 17.0 / 3.0    & 19.1 / 3.8  \\
2       & H$_2^+$                &  & 1.3 / 1.5      & 1.3 / 3.0   & 1.4 / 2.9     & 1.3 / 1.1  \\
3       & H$_3^+$                &  & 0.5 / 0.98     & 0.4 / 2.4   & 0.5 / 2.0     & 0.6 / 0.18 \\
12      & C$^+$                  &  & 1.9 / 0.61     & 2.9 / 0.98  & 3.0 / 0.70    & 4.9 / 0.50  \\
13      & CH$^+$                 &  & 2.8 / 0.37     & 2.8 / 0.78  & 2.8 / 0.78    & 4.9 / 0.78  \\
14      & CH$_2^+$               &  & 4.6 / 0.43     & 4.8 / 0.35  & 4.9 / 0.35    & 5.1 / 0.34 \\
15      & CH$_3^+$               &  & 10.1 / 0.14    & 9.9 / 0.14  & 9.9 / 0.15    & 8.3 / 0.19  \\
16      & O$^+$                  &  & 1.4 / 1.2      & 1.4 / 2.2   & 1.6 / 1.59    & 2.4 / 1.1 \\
17      & OH$^+$                 &  & 2.9 / 0.69     & 2.2 / 0.28  & 2.6 / 0.98    & 3.3 / 1.7 \\
18      & H$_2$O$^+$             &  & 1.6 / 0.05     & 1.5 / 0.03  & 1.6 / 0.04    & 1.1 / 0.12  \\
28      & CO$^+$                 &  & 3.1 / 0.14     & 3.8 / 0.05  & 3.8 / 0.10    & 4.4 / 0.21 \\
29      & COH$^+$ or HCO$^+$     &  & 20.0 / 0.13    & 18.6 / 0.10 & 18.3 / 0.11   & 17.9 / 0.13 \\
30      & HCOH$^+$ or H$_2$CO$^+$ &  & 2.3 / 0.05     & 2.4 / 0.07  & 2.4  / 0.07   & 2.0 / 0.07 \\
31      & H$_2$COH$^+$           &  & 17.3 / 0.04    & 16.2 / 0.05  & 15.9 / 0.06  & 13.8 / 0.07  \\
32      & CH$_3$OH$^+$           &  & 15.8 / 0.03    & 13.9 / 0.03  & 13.8 / 0.03  & 10.8 / 0.03 \\
\hline \hline
\end{tabular}
\end{table*}

We have determined the kinetic energy of all cationic fragments from the X-ray
photodissociation of methanol. The present time-of flight spectrometer was designed to
fulfil the Wiley-McLaren conditions for space focusing (Wiley \& McLaren 1955). Within
the space focusing conditions, the observed broadening of peaks in spectra is mainly due
to the spread in kinetic energy of the fragments released. Considering that the electric
field in the interaction region is uniform, we can determine the released energy in the
fragmentation process ($U_0$) from each peak width, with the formula used by Hansen et
al. (1998) and Santos, Lucas \& de Souza (2001)

\begin{equation} \label{eq-U0}
U_0 = \Big(\frac{qE \Delta t}{2} \Big)^2 \frac{1}{2m}
\end{equation}
where $q$ is the ion fragment charge, $E$ the electric field in the
interaction region, $m$ is the mass of the fragment, and $\Delta t$
is the time peak width (FWHM) taken from PEPICO spectra. In order to
test the above equation we have measured the argon mass spectrum
under the same conditions.

The calculated values for kinetic energy release ($U_0$) for methanol fragmentation are
shown in Table~\ref{tab:piy}. We observe that the highest kinetic energy release (up to
3.8 eV) was associated with the lightest fragment H$^+$ ($m/q=1$) followed by $H_2^+$
($m/q=2$), $H_3^+$ ($m/q=3$) and other ions like C$^+$ and O$^+$, as expected. In our
previous formic acid photodissociation results (PAPER I), we found extremely fast ionic
fragments ($U_0> 10$ eV) that were associated with dissociation of doubly or
multiply-charged ions at high photon energies.

The surface potentials of the ionic states are extremely repulsive. For core excited
molecules which dissociate into one charged and one or more neutral fragments, the
dissociation is primarily controlled by chemical (non-Coulomb) forces originating from
the residual valence electrons of the system (Nenner \& Morin, 1996). From
Table~\ref{tab:piy}, one can see that the mean kinetic energy release, $U_0$, of some
methanol fragments increases as the photon energy approximates the C 1s edge (288-292
eV). This enhancement is due to the repulsive character of the $\sigma$* ($\pi$*)
resonance.

\subsection{Photodissociation and formation pathways}

The present work shows the photofragmentation of methanol produced by soft X-rays and
compares the yields of these fragments with those produced by 70 eV electrons. The inner
shell photoionization process may produce instabilities on molecular structure (nuclear
rearrangements) leading them to peculiar dissociation pathways. From our data we could
determine the main photodissociation pathways from single ionization over the C1s photon
energy range. These photodissociation pathways are shown in Table~\ref{tab-path1} for
288 eV. Only events greater than 2\% were considered here. The main dissociation leads
to production of COH$^+$ + H$_2$ + H ($\sim$ 19 \%). However the single C-H bond rupture
as a result of both H$^+$ + neutrals fragments or H$_2$COH$^+$ + H, represent a combined
yield of about 34 \% of the photodissociation channels. The fragments released due to
the C-O bond rupture also represent an important route of photodissociation ($\sim$ 12
\%), for example the CH$_3^+$ + OH and CH$_3$ + OH$^+$ product routes.

Herbst \& Leung (1986) have presented several pathways for the synthesis of complex
molecules in dense interstellar clouds via gas-phase chemistry models. The authors
presented a significant amount of normal ion-molecule reactions including the ions
C$^+$, CH$^+$, OH$^+$, CO$^+$, CH$_2^+$, H$_2$O$^+$, HCO$^+$, CH$_3^+$, etc.  In another
set of reactions they have shown several radiative association reactions including the
ions C$^+$, CH$_3^+$, HCO$^+$, could lead to the production of high molecular complexity
species. This work points out the importance of the ionic species in the increase of
interstellar molecular complexity.

As a consequence of the high production of COH$^+$ (or HCO$^+$) from methanol
photodissociation by soft X-rays, we expect that a fraction of the detected COH$^+$ (or
HCO$^+$) in the interstellar medium (mainly at XDRs and HMCs) comes from methanol, since
CH$_3$OH is one of the mostly abundant molecule in these regions.

The work presented here strongly suggests that a great number of ions could be produced
by the X-ray photodissociation of large organic molecules in star forming regions.
Therefore, the knowledge of the photodissociation processes and its ions yields has an
essential role in the interstellar chemistry scenario.

\begin{table}[!t]
\caption{Main photodissociation pathways from single ionization due to soft X-ray
photons (288 eV)} \label{tab-path1}
\begin{center}
\setlength{\tabcolsep}{4pt}
\begin{tabular}{l c l}
\hline \hline
CH$_3$OH + $h\nu$ & $\longrightarrow$ & CH$_3$OH$^{+} + e^-$    \\
\hline
CH$_3$OH$^+$    & $\stackrel{18.6\%}{\longrightarrow}$   & COH$^+$ + 3 H (or H$_2$ + H)  \\
                & $\stackrel{17.3\%}{\longrightarrow}$ & H$^+$ + neutrals \\
                & $\stackrel{16.2\%}{\longrightarrow}$   & H$_2$COH$^+$ + H \\
                & $\stackrel{9.9\%}{\longrightarrow}$   & CH$_3^+$ + OH (or O$_2$ + H) \\
                & $\stackrel{4.8\%}{\longrightarrow}$    &  CH$_2^+$ + H$_2$O (or OH$_2$ + H) \\
                & $\stackrel{3.8\%}{\longrightarrow}$  & CO$^+$ + neutrals \\
                & $\stackrel{2.8\%}{\longrightarrow}$    & CH$^+$ + neutrals \\
                & $\stackrel{2.9\%}{\longrightarrow}$  & C$^+$ + neutrals \\
                & $\stackrel{2.4\%}{\longrightarrow}$  & HCOH$^+$ + H$_2$ (or H + H) \\
                & $\stackrel{2.2\%}{\longrightarrow}$  & OH$^+$ + CH$_3$ (or CH$_2$ + H) \\
\hline \hline
\end{tabular}
\end{center}
\end{table}

\subsection{Absolute photoionization and photodissociation cross sections}

The absolute cross section values for both photoionization ($\sigma_{ph-i}$) and
photodissociation ($\sigma_{ph-d}$) of organic molecules are extremely important as
input for molecular abundances models. Sorrell (2001) has presented a theoretical model
in which biomolecules are formed inside the bulk of icy grain mantles photoprocessed by
starlight (ultraviolet and soft X-rays photons). However, the main uncertainty of this
equilibrium abundance model comes from the uncertainty of the $\sigma_{ph-d}$ value.
Therefore the precise determination of $\sigma_{ph-d}$ of biomolecules is very important
to properly estimate the molecular abundance of those molecules in the interstellar
medium that have been produced by this mechanism. Moreover, knowing the photon dose
$I_0$ and $\sigma_{ph-d}$ values, it is possible to determine the half-life of a given
molecule, as discussed by Bernstein et al. (2004).

\begin{figure}[!tb]
\resizebox{\hsize}{!}{\includegraphics{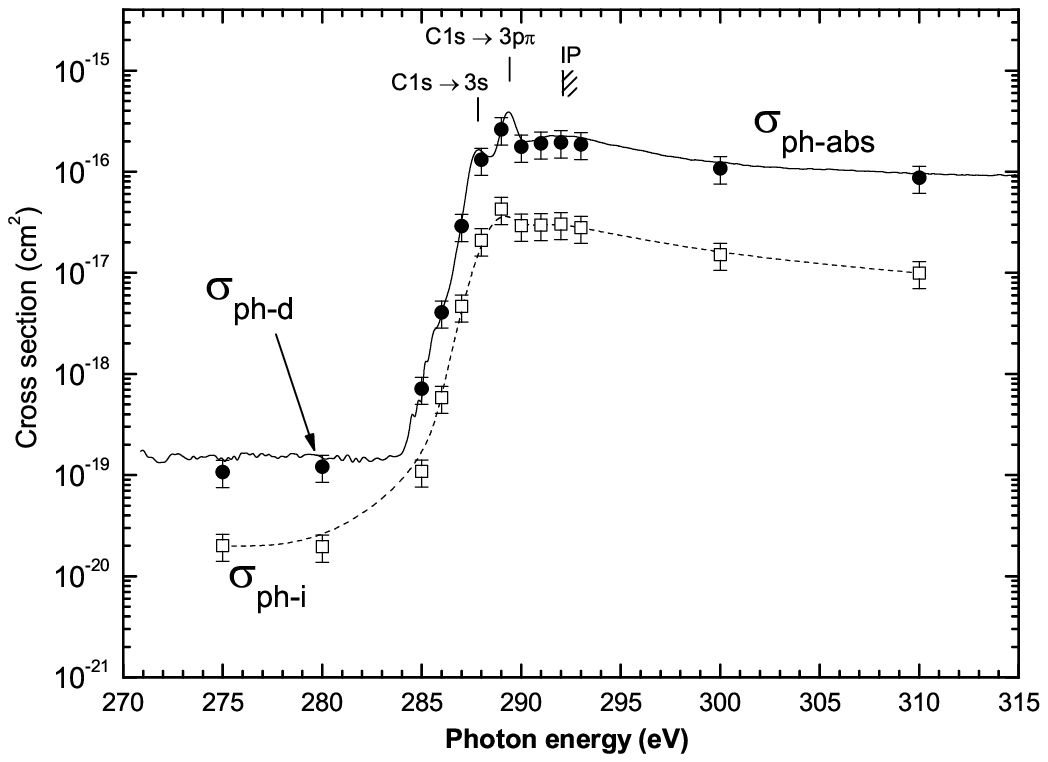}} \caption{Non-dissociative single
ionization (photoionization) cross section, $\sigma_{ph-i}$
(\tiny$\square$\footnotesize), and dissociative ionization (photodissociation) cross
section, $\sigma_{ph-d}$ ($\bullet$), of methanol as a function of photon energy. The
photoabsorption cross-section, $\sigma_{ph-abs}$ (solid line), taken from Ishii \&
Hitchcook (1988) is also shown.}
 \label{fig-sigma}
\end{figure}

\begin{table}[!htb]
\centering %
\caption{Values of non-dissociative single ionization (photoionization) cross section,
$\sigma_{ph-i}$, and dissociative ionization (photodissociation) cross section,
$\sigma_{ph-d}$, of CH$_3$OH as a function of photon energy. The estimated experimental
error was 30\%. The photoabsorption cross section ($\sigma_{ph-abs}$) from Ishii \&
Hitchcook (1988) is also shown.} \label{tab-sigma}
\begin{tabular}{ l l c c c }
\\
\hline \hline
Photon       &  & \multicolumn{3}{c}{Cross Sections (cm$^{2}$)}\\
\cline{3-5}  energy (eV)  &  & $\sigma_{ph-d}$ & $\sigma_{ph-i}$    & $\sigma_{ph-abs}$ \\
\hline
275         &   & 1.1$\times 10^{-19}$ &  2.0$\times 10^{-20}$  & 1.3$\times 10^{-19}$ \\
280         &   & 1.2$\times 10^{-19}$ &  1.9$\times 10^{-20}$  & 1.4$\times 10^{-19}$ \\
285         &   & 7.1$\times 10^{-19}$ &  1.1$\times 10^{-19}$  & 8.2$\times 10^{-19}$ \\
286         &   & 4.1$\times 10^{-18}$ &  5.8$\times 10^{-19}$  & 4.6$\times 10^{-18}$ \\
287         &   & 2.9$\times 10^{-17}$ &  4.6$\times 10^{-18}$  & 3.4$\times 10^{-17}$ \\
288         &   & 1.3$\times 10^{-16}$ &  2.1$\times 10^{-17}$  & 1.5$\times 10^{-16}$ \\
289         &   & 2.6$\times 10^{-16}$ &  4.3$\times 10^{-17}$  & 3.1$\times 10^{-16}$ \\
290         &   & 1.8$\times 10^{-16}$ &  2.9$\times 10^{-17}$  & 2.1$\times 10^{-16}$ \\
291         &   & 1.9$\times 10^{-16}$ &  2.9$\times 10^{-17}$  & 2.2$\times 10^{-16}$ \\
292         &   & 1.9$\times 10^{-16}$ &  3.0$\times 10^{-17}$  & 2.1$\times 10^{-16}$ \\
293         &   & 1.9$\times 10^{-16}$ &  2.8$\times 10^{-17}$  & 2.2$\times 10^{-16}$ \\
300         &   & 1.1$\times 10^{-16}$ &  1.5$\times 10^{-17}$  & 1.2$\times 10^{-16}$ \\
310         &   & 8.7$\times 10^{-17}$ &  9.9$\times 10^{-18}$  & 9.7$\times 10^{-17}$ \\
\hline \hline
\end{tabular}
\end{table}

In order to put our data on an absolute scale, after subtraction of a linear background
and false coincidences coming from aborted double and triple ionization (see Simon et
al. 1991), we have summed up the contributions of all cationic fragments detected and
normalized them to the photoabsorption cross sections measured by Ishii \& Hitchcook
(1988). Assuming a negligible fluorescence yield (due to the low carbon atomic number
(Chen et al. 1981)) and anionic fragments production in the present photon energy range,
we assumed that all absorbed photons lead to cation formation.

The absolute cross section determination is described elsewhere (PAPER I, PAPER II).
Briefly, the non-dissociative single ionization (photoionization) cross section
$\sigma_{ph-i}$ and the dissociative single ionization (photodissociation) cross section
$\sigma_{ph-d}$ of methanol can be determined by
\begin{equation}
\sigma_{ph-i} = \sigma^{+} \frac{PIY_{CH_3OH^+}}{100}
\end{equation}
and
\begin{equation}
\sigma_{ph-d} = \sigma^{+} \Big( 1 - \frac{PIY_{CH_3OH^+}}{100}
\Big)
\end{equation}
where $\sigma^{+}$ is the cross section for single ionized fragments
(see description in PAPER I and PAPER II).

Both cross sections can be seen in Figure~\ref{fig-sigma} as a function of photon
energy. The absolute absorption cross section of methanol (Ishii \& Hitchcook 1988) is
also shown for comparison. Those values are also listed in Table~\ref{tab-sigma}. The
estimated experimental error was about 30\% which includes the data treatment, the
uncertainties in the gas density in the interaction region and the loss of extremely
energetic ions (initial kinetic energy $>$ 30 eV).

\section{Summary and conclusions}

The aim of this work was to experimentally study the ionization and dissociation
processes of the simplest interstellar alcohol, CH$_3$OH, employing soft X-ray photons
(100-310 eV) that go out from a toroidal grating monochromator (TGM) beamline of the
Brazilian Synchrotron Light Laboratory (LNLS). The experimental set-up consists of a
high vacuum chamber with a time-of-flight mass spectrometer (TOF-MS). Mass spectra were
obtained using coincidence techniques.

Several ionic fragments produced by methanol photodissociation, have considerable
kinetic energy (e.g. H$^+$, H$_2^+$, C$^+$ and O$^+$). An extension of this process to
conditions in the interstellar medium suggests endothermic ion-molecule (or
radical-molecule) reactions may become important in elucidating the pathways of
formation of some complex molecules (Largo et al. 2004). Unlike previous work with
formic acid (PAPER I) performed in the same spectral range, no fragments with large
kinetic energy fragments have been observed due the single photoionization of CH$_3$OH.

We have found that about 11-16\% of CH$_3$OH survives soft X-ray ionization field.
COH$^+$ (or HCO$^+$), H$^+$ and CH$_2$OH$^+$ were the main fragments produced by high
energy photons. This large resistance to the X-ray photons could help to sustain the
large column density observed for methanol in star-forming regions.

Dissociative and non-dissociative photoionization cross sections in
the energy range of 275 eV to 310 eV (over the C1s edge) were also
determined. We hope that the molecular cross sections derived in
this work will give rise to more precise values for some molecular
abundances in interstellar medium chemistry models.

%
\begin{acknowledgements} The authors would like to thank the staff
of the Brazilian Synchrotron Facility (LNLS) for their valuable help
during the course of the experiments. We are particularly grateful
to Dr. R. L. Cavasso and Professor A. N. de Brito for the use of the
Time-of-Flight Mass Spectrometer. This work was supported by LNLS,
CNPq and FAPERJ.
\end{acknowledgements}
%
%

\end{document}